# Temperature and misorientation-dependent austenite nucleation at ferrite grain boundaries in a medium manganese steel: role of misorientation-dependent grain boundary segregation

*Rama Srinivas Varanasi[1,2], #Osamu Waseda[1], Faisal Waqar Syed[1], Prithiv Thoudden-Sukumar[1], Baptiste Gault[1,3], Jörg Neugebauer[1], †Dirk Ponge[1]

[1]Max-Planck-Institut for Sustainable Materials, Max-Planck-Straβe 1, 40237, Düsseldorf, Germany

[2] Institute for Materials Research, Tohoku University, 2-1-1 Katahira, Aoba-ku, Sendai, 980-8577 Japan

[3]Department of Materials, Royal School of Mines, Imperial College, Prince Consort Road, London SW7 2BP, United Kingdom.

**Corresponding authors**

*Rama Srinivas Varanasi: rama.varanasi@tohoku.ac.jp
#Osamu Waseda: o.waseda@mpie.de
†Dirk Ponge: d.ponge@mpie.de

## Graphical abstract

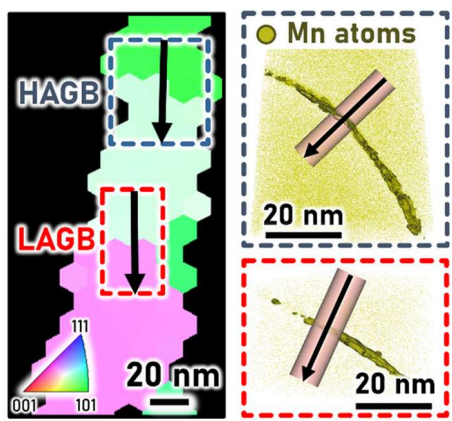
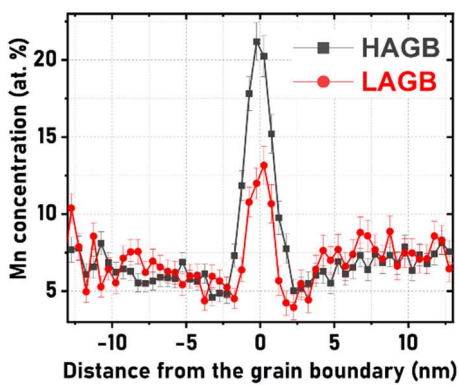
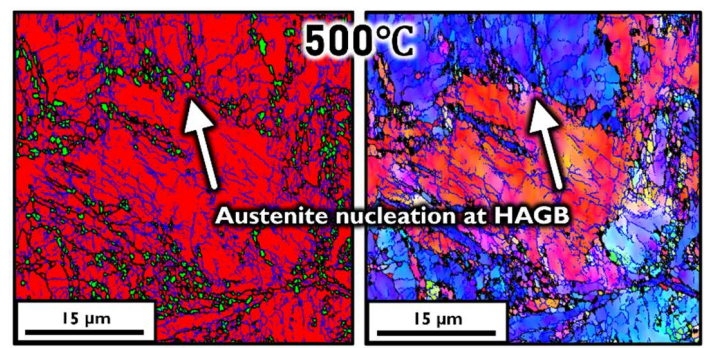
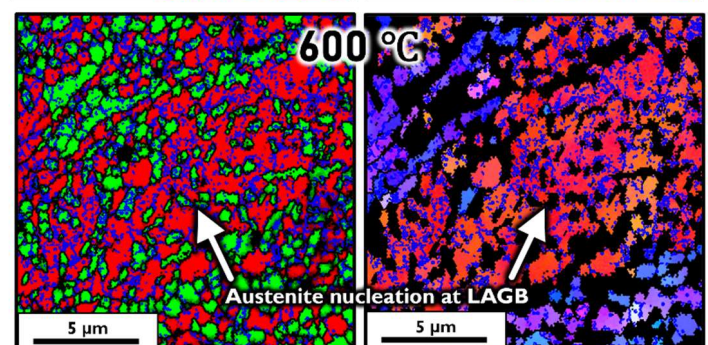
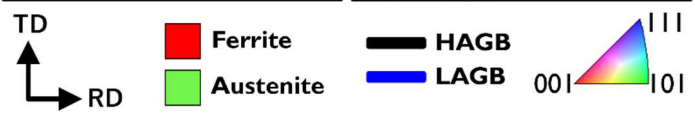

I

## Abstract

In the current work, we study the role of grain boundary (GB) misorientation-dependent segregation on austenite nucleation in a 50% cold rolled intercritically annealed 10Mn-0.05C-1.5Al (wt. %) medium Mn steel. During intercritical annealing at 500°C, austenite nucleates predominantly at high-angle GBs. At 600°C, austenite nucleates additionally at low-angle GBs, exhibiting a temperature dependance. Correlative transmission Kikuchi diffraction /atom probe tomography reveals a misorientation-dependent segregation. While GB segregation has been reported to assist austenite nucleation in medium manganese steels (3-12 wt.% Mn), an understanding of the temperature and misorientation dependance is lacking, which is the aim of current work. Since artifacts of atom probe can cause a broadening of the segregation width, we combined experiments with results from density functional theory (DFT) calculations that reveal that the Mn segregation is not limited to the GB plane but confined to a region in the range of approximately 1 nm. Consequently, GB segregation alters both the GB interface energy and the free energy per unit volume corresponding to the transformation. We estimate the local driving force for austenite nucleation accounting for the segregation width of ~ 1 nm. Based on classical nucleation theory, we clarify the effect of GB segregation on the critical radius and activation energy barrier for confined austenite nucleation at the GB.

## I. Introduction

Ultrafine-grained ferrite-austenite dual-phase medium manganese steels (3-12 wt.% Mn) are promising candidates as structural materials for automotive parts due to their good balance between production costs and mechanical properties [1,2]. The ferrite-austenite dual-phase microstructure in medium Mn steels is obtained by austenite reversion from martensite during intercritical annealing [3]. Nakada et al. [3] have attributed the rapid austenite formation from the martensitic matrix to two factors, a higher density of nucleation sites in the martensitic structure and higher diffusivity of Mn in the martensite. Recently, mechanisms of the defect-driven (dislocations and grain boundaries) austenite growth during the $\alpha' \rightarrow \gamma$ transformation have been discussed [4,5]. However, the role of GB misorientation-driven segregation in austenite nucleation is still unclear.

Typically, solid-state phase transformations occur via heterogeneous nucleation at defects such as grain or phase boundaries, triple junctions, etc. It is caused by three microstructural factors: (i) Interface energy of these defects [6]; (ii) Second-order defects along/on interfaces (such as ledges and steps) [7]; (iii) Stored defect energy at the grain boundaries (GBs) [8,9]. These three factors are further affected by the solute segregation to the defects. In intercritically annealed medium Mn steels ,recent studies [10–15] have revealed Mn segregation to defects, including dislocations and GBs, as a pathway for austenite nucleation. While the role of GBs as austenite nucleation sites is uncontested, it was reported that linear complexions formed due to Mn segregation at the dislocations are subcritical nuclei of austenite [16]. GB spinodal segregation has been reported to be a precursor state for the heterogeneous nucleation of austenite at GBs [12–14], and models were proposed to predict GB segregation in solid solutions [13,14]. While the effect of GB misorientation on solute segregation has previously been discussed in the literature [17–22] for various alloy systems, such a study regarding the role of GB misorientation on Mn segregation is lacking in medium Mn steels. Furthermore, understanding of the effect of GB misorientation-dependent segregation on austenite nucleation is lacking, which is the aim of the present work.

The greater the initial GB energy ($\Gamma_{\alpha\alpha}$), the higher the driving force for phase transformation, and the lower the activation energy barrier ($\Delta G^*_{het}$) when the GB is eliminated during heterogeneous nucleation (Supplementary S1). The solute segregation to GBs is driven by the reduction in free energy. Thus, GB segregation decreases the free energy available for release during the elimination of the GB during the nucleation of a new phase. Consequently, GB segregation should result in the reduction of the driving force for nucleation and an increase in the activation energy barrier. Despite that, GB segregation is reported to promote heterogeneous nucleation [10,13]. From the perspective of classical nucleation theory, this appears paradoxical. The present work aims to resolve this apparent contradiction by estimating the effect of GB segregation on the $\Delta G^*_{het}$.

We study the role of GB misorientation on Mn segregation and the subsequent role of Mn segregation in austenite nucleation in a 10Mn-0.05C-1.5Al (wt. %) medium Mn steel. To this end, we employ a correlative transmission Kikuchi diffraction (TKD)/atom probe tomography (APT) approach. APT's resolution is highest in specific regions within the dataset, and not across the entire reconstruction [23]. In the case of pure metals, while the depth resolution is <0.1 nm, the lateral resolution is <1 nm [24] at best. In laser-pulsing mode, the lateral resolution is further degraded by lateral atomic displacements by thermally activated surface diffusion processes during the pulsed-laser mode [25,26]. The spatial resolution further degrades at an interface with solute segregation [27–29], because of the difference in the field evaporation behaviour associated to the change in local GB chemistry and atomic density [27–29]. Therefore, the GB segregation data obtained by APT measurements must be accounted for, for such inaccuracies. Besides, the GB segregation need not be limited to only the GB plane [30–34]. Hence, it is necessary to interpret the GB segregation values determined from APT with caution. To this end, in the current work, we benchmark the GB segregation width obtained from APT based on density functional theory (DFT) calculations. We demonstrate that the GB segregation is confined to ~1 nm, and it thus alters the local chemical driving force for austenite nucleation. The effect of GB segregation on the critical radius (r*) and activation energy barrier ($\Delta G^*_{het}$) for austenite nucleation is clarified based on classical nucleation theory. The insights gained from this study can be extended to understand the GB segregation-assisted nucleation in other alloy systems.

## 2. Material and methods

### 2.1. Alloy synthesis and heat treatment

The alloy 10Mn-0.05C-1.5Al (wt. %) was hot rolled, solution-annealed for 3 hours at 1100°C, and later quenched. Subsequently, the alloys were cold rolled (50% thickness reduction). The samples were then intercritically annealed in an argon atmosphere in a quartz tube furnace for 6 hours. A 10°C/min heating rate was employed, and the samples were furnace-cooled after annealing. We abbreviate the intercritically annealed samples as IAT (where T is the intercritical annealing temperature in °C).

In order to only study the role of GB segregation on austenite nucleation, it is essential to avoid carbide formation. Thermocalc calculations showed no cementite formation for intercritical annealing temperatures greater than 500°C [4]. Thus, in the present work, we investigate the samples intercritically annealed at 500°C. Further, to study the effect of intercritical annealing temperature on the austenite nucleation, we also investigate the sample intercritically annealed at 600°C.

## 2.2. Material characterization

For electron backscatter diffraction (EBSD) investigations, the samples were ground from 220 grit up to 1000 grit carbide sandpaper, followed by 3 μm diamond paste polishing. Subsequently, the oxide polishing suspension (OPS) polishing was carried out with water-free 50 nm colloidal silica suspension. EBSD studies were carried out in a JEOL JSM 6500 field emission scanning electron microscope (SEM) with an EDAX Digiview IV EBSD detector. The OIM Analysis 7 software was used for post-processing the EBSD scans. In the present work, we refer to GBs with misorientation of 2°-15° as low-angle grain boundaries (LAGBs), and GBs with misorientation angle of 15°-63° as high-angle grain boundaries (HAGBs). Only the indexed points with confidence index (CI) values greater than 0.1 are considered in the present work.

In the current work, we employ a correlative approach consisting of transmission Kikuchi diffraction (TKD) and atom probe tomography (APT) to obtain the structural and chemical information of the ferrite GB. Atom probe specimen preparation and TKD investigations (CI > 0.1) on the atom probe needles were carried out in a dual-beam FEI Helios NanoLab 660i scanning-electron microscope (SEM) - focused ion beam (FIB) coupled with an EDAX Hikari Plus EBSD camera. A Standard lift-out protocol was employed for sample preparation [35]. We carried out the APT investigation at 60 K in a laser pulsing mode with a repetition rate of 125 kHz and 40 pJ pulse energy using a LEAP 5000XR HR with a reflectron (Cameca instrument). Cameca's IVAS (3.8.4) and AP suite software were used for APT reconstructions.

## 2.3. Density functional theory (DFT) calculations

In the *ab initio* calculations, we consider 8 different types of coincidence site lattice (CSL) grain boundaries (Table 1). To find the global energy minimum, we calculated the grain boundary of each of the simulation box using LAMMPS with the Ackland potential [36] (in order to pre-screen structures) by varying the following: (1) box length along the direction perpendicular to GB; (2) number of atomic layers on the GB plane; (3) gamma surface of GB. For each of the boxes, a structure optimization with conjugate gradient was performed. Then the minimum energy structure is used for further study with DFT. Fe atoms were set to a ferromagnetic state and then allowed to converge to the minimum energy. For Mn atoms, we took all possible ferromagnetic and anti-ferromagnetic alignments, as long as the number of possible states was smaller than 10, to find the minimum energy value. With more than 10 possible states, we took 10 states randomly and took the configuration with the lowest final energy. With this strategy, we might have missed the global minimum, but as an earlier study shows the energy variation due to the variation of Mn magnetic states is relatively small [37].

Given that we do not observe segregation of Al in the APT results, we did not include Al in the simulations. For the sake of simplicity, we do not consider C in the present calculations. Previous studies have shown that the Mn segregation is enhanced through the presence of C [38,39]. Therefore, we can consider our results to be the lower limit of the segregation energy.

*Table 1 Summary of the GBs investigated.*

| GB geometry | GB type | Misorientation (°) |
|---|---|---|
| Σ3[110](110) | Twist | 70.53 |
| Σ3[111](111) | Twist | 60 |
| Σ3[211](211) | Twist | 180 |
| Σ5[100](012) | Tilt | 53.13 |
| Σ5[310](310) | Twist | 180 |
| Σ9[110]($\bar{1}1\bar{4}$) | Tilt | 38.94 |
| Σ13[320](320) | Twist | 180 |
| Σ25[100](0$\bar{4}$3) | Tilt | 73.74 |

In this study, we present a comparison between single segregation, i.e., the segregation of a single Mn, and collective segregation, i.e., segregation of multiple Mn atoms, in which the interactions between Mn atoms are also considered. Detailed rationale will be provided in section 3.3. To calculate the collective segregation of Mn atoms, the simulations were performed following these two steps:

i. In each simulation, an Fe atom was replaced by a Mn atom and then the segregation energy was calculated.

ii. Following the Langmuir–McLean theory [40], we estimated the occupation fraction (at annealing temperature of 500 °C and base concentration of 10 at. % Mn) using the segregation energies we obtained in step 1 and distributed Mn atoms accordingly. In order to calculate the chemical potential of Mn, special quasi-random structures of Fe-Mn with a total of 53 atoms were created. Then the energy values were interpolated by a linear regression.

In order to account for many body chemical interactions, we introduce the occupation factor κ ∈ [0,1] to approximate the effective segregation energy $E_i^{eff}$ for the site $i$ via:

$$E_i^{eff} = \kappa\left(E_i + p(E_i^{eff})E_{chem}\right) \qquad 1$$

where $p(E)$ is the occupation fraction following the Fermi-Dirac statistics and $E_{chem}$ is the effective chemical repulsion energy for the system, which is fitted by interpolating the individual and collective segregation energies, so that for κ = 0, the individual segregation energies are reproduced, and for κ = 1, the sum of $E_i^{eff}$ reproduces the collective segregation energies. The value of κ is chosen in the way that the total energy becomes the minimum, i.e.

$$\kappa = min_\kappa \left\{ \sum_i p(E_i^{eff}) E_i^{eff} \right\} \qquad 2$$

## 2.4. Thermodynamic calculations

Thermodynamic information such as phase fraction, chemical composition, and Gibbs free energy of the phase(s) at different intercritical annealing temperatures was determined by the CALPHAD approach employing TCFE9 database in Thermocalc (2017b).

## 2.5. Grain boundary excess calculations

The APT GB segregation width profiles vary for different measurements. However, this could arise due to the inherent segregation behavior of the material [30] and/or also related to APT-related evaporation artifacts. These artefacts are caused by differences in the field evaporation behavior of atoms located at the grain boundary itself [28], because of the different atomic configurations only, or the local differences in composition [27]. The relative orientation of the grain boundary with respect to the specimen's main axis can also play a role, since APT's resolution is not isotropic [29].

To address this limitation in the present work, we calculate the interfacial excess values of the solutes at the GB. This gives an insight into the total inherent segregation associated with the GB, thereby partly accounting for inaccuracies in segregation width profiles arising due to APT-related artifacts. Earlier, Krakauer et al. [41] and Maugis et al. [42] proposed different methodologies to calculate the interfacial excess of solute at a GB. The method proposed by Krakauer et al. [41] attributes the excess solute at the interface only to the GB plane. However, the solute segregation need not be limited only to the GB plane [30] as the solute can also enrich the bulk lattice adjacent to the GB plane. Therefore we employ the method reported by Maugis et al. [42] (explained in detail in supplementary S2). This approach will be further justified in section 3.3 based on *ab initio* calculations. Maugis et al. [42] proposed that the integral GB excess values for the measured values should be similar to that of the actual segregation. The integral GB excess is represented in the units at. % nm. In other words, it is the GB excess associated with a uniform GB segregation across 1 nm. In the present work, we refer to this as 'adjusted GB segregation value' and it is used to calculate the driving force for the phase transformation (section 3.3). The rationale behind considering the GB width as 1 nm is explained in section 3.3.

# 3. Results
## 3.1. Microstructure

Figs. 1a and 1c depict the phase maps of the IA500 and IA600, respectively (intercritically annealed for 6 hours). Figs. 1b and 1d illustrate the corresponding inverse pole figure (IPF) maps for the IA500 and IA600, respectively. The phase fraction of austenite measurement from EBSD studies is ~0.1 and ~0.4 in the IA500 and IA600, respectively. Table 2 compares the measured austenite phase fraction to that of Thermo-Calc calculations. It can be noted that while the measured austenite phase fraction for the IA500 is one-third of the estimated austenite phase fraction, the measured austenite fraction is closer to the Thermo-Calc prediction for the IA600. In Figs. 1a-b, for the IA500, austenite nucleation is predominantly observed at the high-angle grain boundaries (HAGBs) as highlighted by the white arrow. However, for the IA600 (Figs. 1c-d), austenite is observed both at the high-angle grain boundaries (HAGBs) and low-angle grain boundaries (LAGBs). This demonstrates temperature-dependent heterogeneous nucleation and growth of austenite.

**Table 2** Austenite phase fraction in the IA500 and IA600 compared with the Thermo-Calc estimations.

|  | IA500 | IA600 |
| --- | --- | --- |
| **Measured** | 0.1 | 0.4 |
| **Thermo-Calc estimation** | 0.3 | 0.47 |

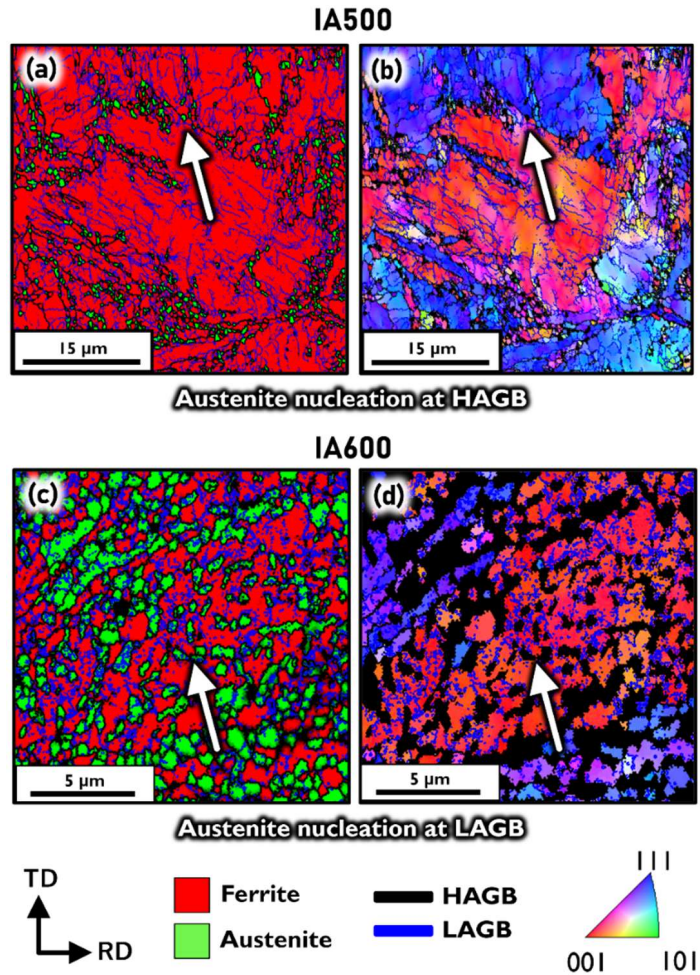

*Figure 1 Heterogeneous austenite nucleation revealed from electron backscattered diffraction (EBSD) studies. (a) Phase map for IA500 (b) Corresponding inverse pole figure (IPF) map of ferrite for IA500 (c) Phase map for IA600 (d) Corresponding IPF map of ferrite for IA600. Low-angle grain boundaries (LAGBs) correspond to GB misorientation of 2°-15°, and high-angle grain boundaries (HAGBs) correspond to GB misorientation angle of 15°-63°*

## 3.2. Thermodynamic Calculations

Gibbs free energy values for austenite and ferrite for varying Mn and C concentrations at 500°C and 600°C are obtained from Thermocalc (Gibbs energy maps shown in Fig. 2). These values are subsequently used to calculate the molar chemical driving force for austenite nucleation ($\Delta G_{chem}$) in sections 3.4 and 4.2, using equation 3. The rationale will be explained in detail in section 4.1.

$$\Delta G_{chem} = G_v^\gamma - G_v^\alpha \qquad 3$$

wherein, $G_v^\gamma$ and $G_v^\alpha$ are the molar Gibbs energy of the austenite and ferrite, respectively.

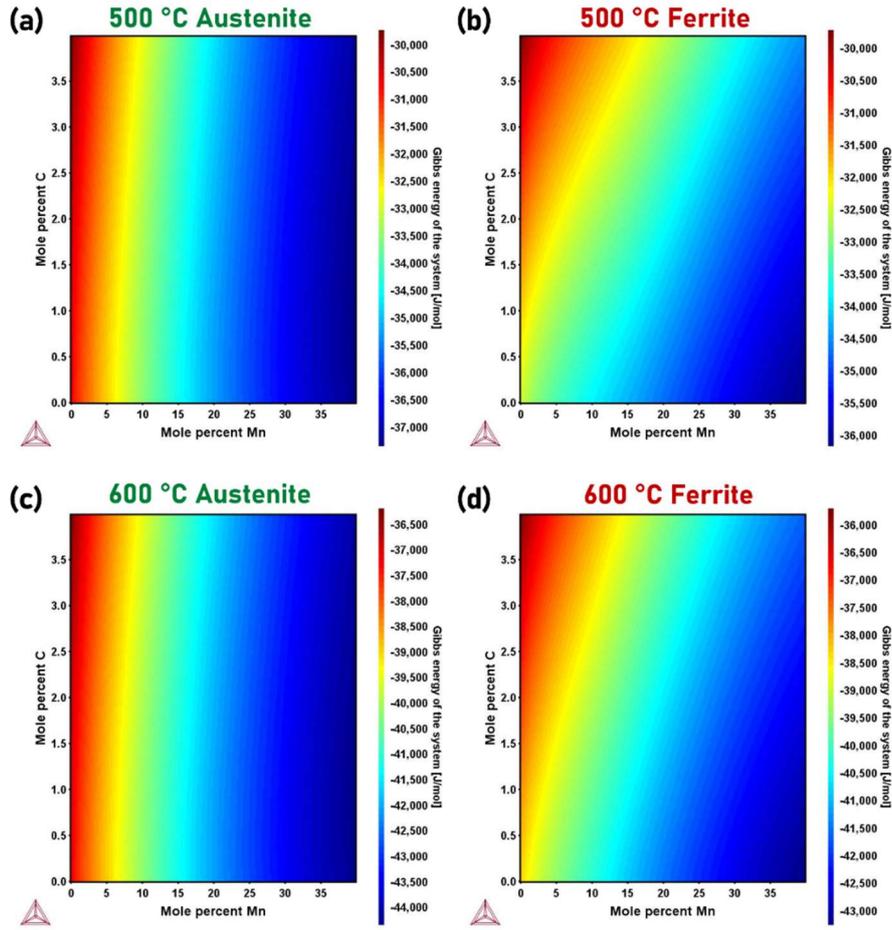

*Figure 2 Gibbs free energy maps for (a) austenite at 500°C (b) ferrite at 500°C (c) austenite at 600°C (d) ferrite at 600°C, obtained using Thermocalc.*

### 3.3. Segregation energies from density functional theory (DFT)

Fig. 3 shows the segregation energies (obtained from DFT calculations at 0 K) of single Mn atoms in an array of selected GBs, for which Mn shows strong segregation tendency. It is important to note that except for the $\sum 3(111)[111]$ and $\sum 9(110)[\bar{1}1\bar{4}]$ GBs, the segregation tendency was maximum in the plane(s) adjacent to the GB plane and not at the grain boundary plane itself. In other words, the segregation is not confined to the GB plane. This observation is in accordance with the DFT calculations by Jin et al. [33], wherein it was observed that for a $\sum 5(113)[100]$ in BCC Fe, the maximum Mn segregation tendency was at a position adjacent to the GB plane. The segregation tendency being highest in the plane(s) adjacent to GB in the Fe-Mn system has magnetic origins [31,43].

For the Mn solute segregation at $\sum 5(113)[100]$ GB in BCC Fe, Jin et al. [33] showed that the segregation energy is negative for the second solute Mn atom segregating at the GB site when a Mn atom is already present at another GB position. Jin et al. [33] explained that the segregation of an additional solute Mn atom remains favorable, with a decrease in segregation tendency. Along similar lines, multiple Mn atoms would segregate at the GB as long as the segregation energy of an additional Mn is negative. During intercritical annealing, the bulk acts as a source of solute Mn atoms, enabling segregation far greater than the base concentration of 10 at. %. Hence, to estimate the GB segregation, we must consider the collective segregation energies of multiple Mn atoms. The detailed methodology was detailed in section 2.3. The *ab initio* computed segregation energies shown in Fig. 3 were used to calculate the occupancy probability for each site following the Langmuir-McLean theory, where we took the base concentration of 10 at. % Mn and the temperature of 500 °C. Based on these results and eqs. (1) and (2), we obtained the Mn fraction (occupation probability) as shown in Fig. 4a.

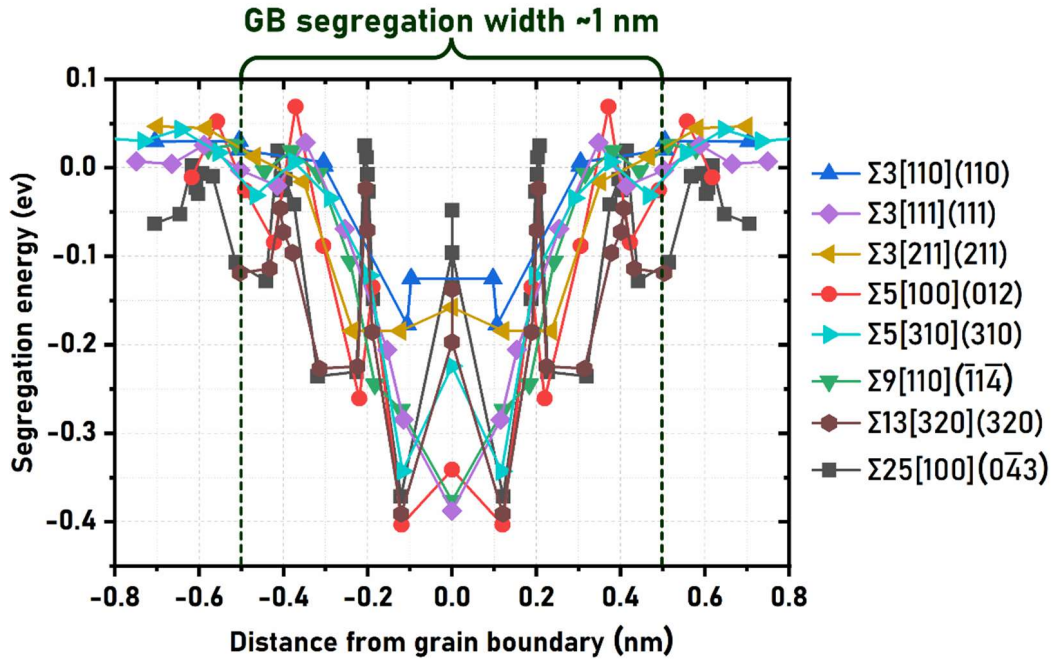

Figure 3 Segregation energies of single Mn in various GB (negative: attractive) are obtained from DFT calculations (0 K).

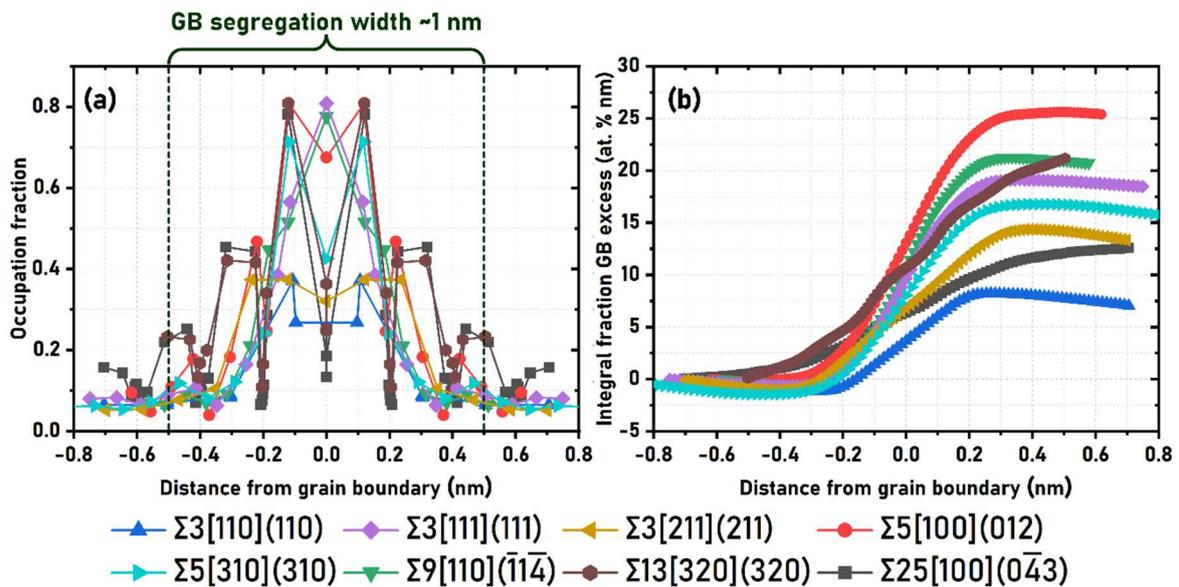

Figure 4 (a) Expected Mn fraction (occupation probability) for collective segregation energies based on the Langmuir-McLean theory (at a base concentration of 10 at. % and a temperature of 500 °C). (b) Integral fraction GB excess of Mn (at. % nm) along the direction perpendicular to the grain boundary.

It can be noted in Figs. 3 and 4a, that the GB segregation thickness is around 0.5 nm on either side of GB plane for nearly all cases. This result is in line with the tracer diffusion experiments that assume the grain boundary width to be in this range [44]. Hence it is essential that we do not attribute the GB solute excess obtained via APT measurements only to the GB plane. Consequently, the approach of Krakauer et al. [41], which assumes the segregation to be limited only to the GB plane is unsuitable for estimating the GB solute excess. Therefore, as mentioned earlier in section 2.5, we estimate the integral fraction of GB excess considering the width of the GB segregation to be 1 nm. Fig. 4b shows the integral fraction GB excess values for all 8 GBs. The change in GB energies following the Mn GB segregation obtained from Langmuir–McLean theory (at 500°C and base Mn concentration of 10 at. %) is summarized in Table 3.

*Table 3 Summary of the GB energies obtained from DFT calculations with the distribution of Mn atoms following the Langmuir–McLean theory at 500°C with the base Mn concentration of 10 at. %.*

| GB geometry | GB energy ($Jm^{-2}$) | | GB energy reduction ratio |
|---|---|---|---|
| | (Before segregation) | (After segregation) | |
| $\Sigma3[110](110)$ | 0.47 | 0.38 | 0.82 |
| $\Sigma3[111](111)$ | 1.48 | 0.88 | 0.60 |
| $\Sigma3[211](211)$ | 0.41 | 0.16 | 0.39 |
| $\Sigma5[100](012)$ | 1.51 | 0.53 | 0.35 |
| $\Sigma5[310](310)$ | 1.43 | 1.05 | 0.73 |
| $\Sigma9[110](\bar{1}1\bar{4})$ | 1.36 | 0.59 | 0.43 |
| $\Sigma13[320](320)$ | 1.30 | 0.59 | 0.45 |
| $\Sigma25[100](0\bar{4}3)$ | 1.20 | 0.90 | 0.75 |
| Average | | | 0.566 |

### 3.4. Correlative transmission Kikuchi diffraction/ atom probe tomography study

To study the effect of GB misorientation on Mn segregation, we employed a correlative TKD/APT approach. The phase map and the corresponding IPF map of the atom probe tip of an IA500 specimen are reported in Figs. 5a-a'. Herein the HAGB and LAGB enclosing the same grain are analyzed. Misorientation profile studies indicate that the HAGB and LAGB have misorientations of 51° and 10°, respectively. The HAGB and LAGB are shown in the inset with 14 at. % and 11 at. % Mn iso-surfaces, respectively. Figs. 5b-b' indicates the one-dimensional (1D) compositional profiles obtained with a 6 nm diameter cylindrical region of interest. In the 1-D solute concentration profiles across GBs, the region with highest Mn concentration at GB is considered the GB (defined as 0 nm).

In accordance with the previous studies [12–14], we observe spinodal decomposition at the GBs; for instance, we observe a Mn compositional spread between 14-21 at. % (Supplementary S3) for the HAGB. However, the driving force for austenite nucleation will be maximal at regions of GB corresponding to higher Mn concentration. Thus, in the current work, we consider the maximum solute concentration. The Mn segregation at HAGB is ~21 at. % and at LAGB is ~13 at. %, as shown in Fig. 5b-b'. We observe that the C and Mn segregation at the LAGB is significantly lower than at the HAGB. To substantiate the lower segregation at LAGB, we further report the GB segregation values for four additional LAGBs ($LAGB_2$ – $LAGB_5$) in table 4.

As seen in Figs. 5b-b', the width of the segregation profiles varies from HAGB to LAGB. We calculated the C and Mn integral GB excess in at. % nm (table 4). The bulk ferrite concentration (shown in table 4 in column '*Mn/C in ferrite at. %*') is estimated by measuring the average concentration (over range of ~5 nm on both sides) 8-10 nm away from the GB. As explained in section 2.5, the adjusted GB solute segregation value (for a GB segregation width of 1 nm) is the sum of its bulk composition and the integral fraction GB excess.

*Table 4 GB excess values for HAGB and LAGBs obtained from APT measurements and the corresponding driving force at 500°C calculated using Thermocalc.*

| GB | Misorien-tation (in °) | Mn Excess (at. % nm) | Mn in ferrite (at. %) | Adjusted GB segregation of Mn (at. %) | C Excess (at. % nm) | C in ferrite (at. %) | Adjusted GB segregation of C (at. %) | $\Delta G_{chem}$ (kJ/mol) |
|---|---|---|---|---|---|---|---|---|
| HAGB | 51 | 23.7 | 7.26 | 30.96 | 3.25 | 0.12 | 3.37 | -1.91 |
| LAGB$_1$ | 10 | 6.05 | 7.09 | 13.14 | 0.27 | 0.14 | 0.41 | -0.47 |
| LAGB$_2$ | 11.4 | 4.8 | 8.79 | 13.59 | 1.12 | 0.16 | 1.28 | -0.79 |
| LAGB$_3$ | 5.5 | 17.23 | 9.15 | 26.38 | 0.02 | 0.14 | 0.16 | -0.76 |
| LAGB$_4$ | 4 | 13.74 | 8.77 | 22.51 | 0.22 | 0.31 | 0.53 | -0.81 |
| LAGB$_5$ | 3.5 | 13.8 | 8.69 | 22.49 | 0.51 | 0.08 | 0.59 | -0.83 |

We consider that the integral fraction of GB excess can be reliably compared across datasets. Thus, we can validate our approach in section 2.3 by comparing the Mn integral fraction of GB excess results obtained from APT and DFT calculations. As seen in table 4, Mn integral fraction GB excess is 23.7 at. % nm for the HAGB. This is comparable to the Mn integral fraction GB excess results obtained from DFT calculation (Fig. 4b) for the ∑5(100)[012] GB (22.5 at. % nm). Similarly, the Mn integral fraction GB excess values of LAGBs in table 4 are comparable to that of other GBs reported in Fig. 4b.

For the HAGB, over the width of 1 nm, the adjusted Mn GB segregation value is 30.96 at. % (Sum of bulk composition and the integral fraction GB excess; table 4). The adjusted Mn GB segregation value is significantly greater than the Mn concentration (~21 at. %) obtained from the 1-D concentration profile in Fig 5b. The width of Mn GB segregation (~4 nm, Fig. 5b) observed in the APT measurement is four times greater than the width of GB segregation calculated using DFT (section 3.3). Therefore, the lower GB segregation measured from the 1-D concentration profile in APT can be attributed to the peak broadening due to the APT-related evaporation artifacts. Furthermore, it also illustrates the merit of calculating the GB integral fraction.

The chemical driving force for austenite nucleation is calculated (as discussed in section 3.2) using the Gibbs energy values corresponding to the adjusted GB segregation values (table 4). It can be observed that the driving force for austenite nucleation at the HAGB is at least twice as much as for a LAGB.

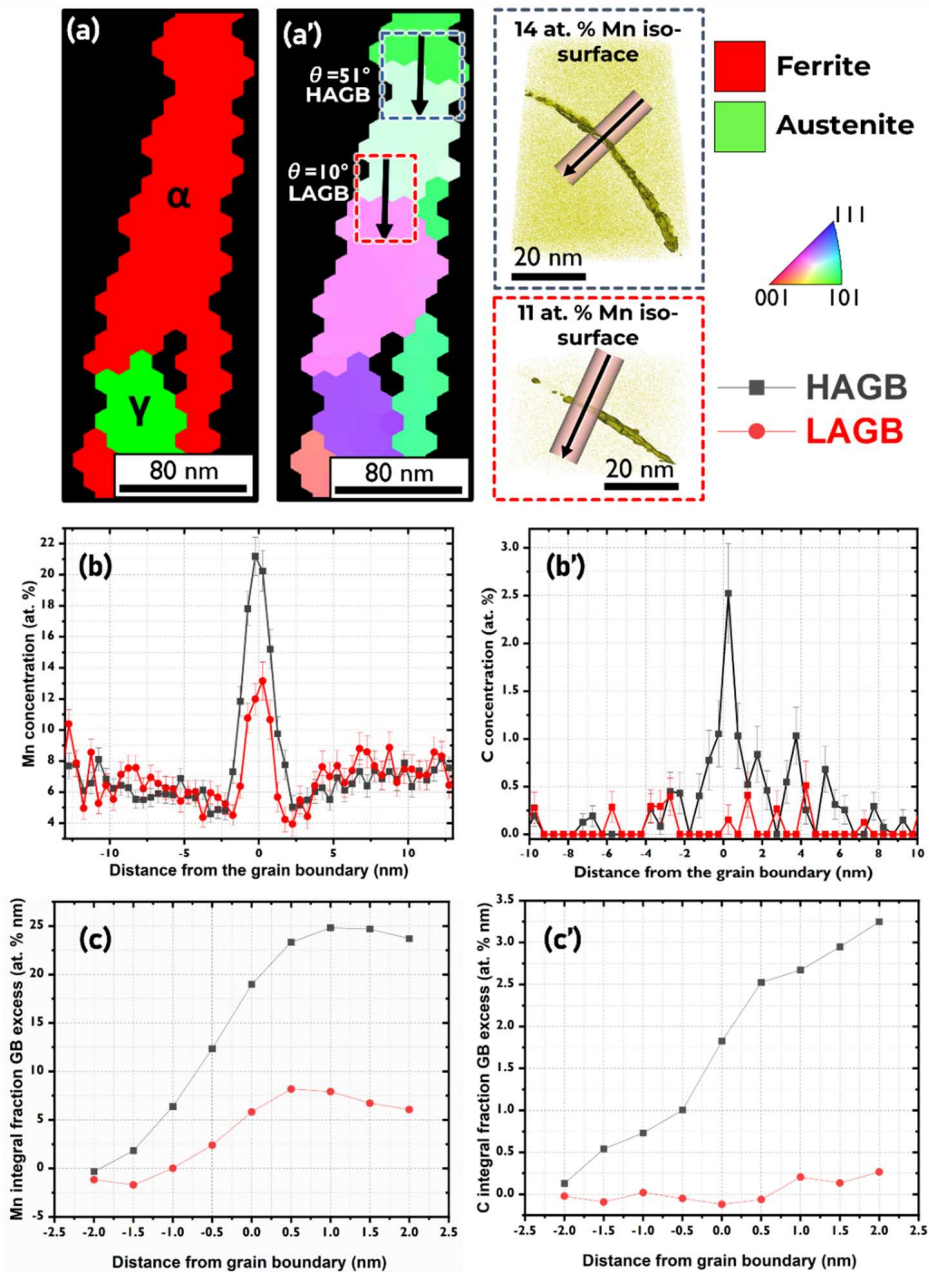

*Figure 5 Correlative TKD/APT study for IA500. TKD studies on an atom probe tip showing (a) phase map and (a') IPF map. High-angle grain boundary (HAGB) highlighted by 14 at. % Mn iso-surface, and (d) low-angle grain boundary (LAGB) highlighted by 11 at. % Mn iso-surface. (b-b') One dimensional (1-D) concentration profiles of Mn and C across the HAGB and LAGB. (c-c') Integral fraction GB excess of Mn and C at the HAGB and LAGB.*

Given the low Mn diffusivity at temperatures lower than 500°C [4], we do not expect any substantial change in the GB chemistry of Mn during furnace cooling for the IA500 (intercritically annealed for 6 hours). However, Mn diffusion to GB can occur during furnace cooling for samples intercritically annealed at 600°C. Furthermore, Mn partitions into austenite from ferrite during intercritical annealing. Therefore, the greater the austenite phase fraction, the lower the ferrite Mn concentration. Since the austenite phase fraction in the IA500 is significantly low (0.1, table 2), there is no significant change in ferrite chemistry (bulk ferrite Mn concentration is 9-10 at. % away from the HAGB in Fig. 5b). On the contrary, for IA600 the austenite phase fraction is 0.4. Consequently, the Mn concentration in ferrite is ~ 4 at. %, as reported elsewhere [45]. In order to study Mn GB segregation, an intercritical annealing temperature of 600°C thus necessitates a shorter annealing time (to avoid austenite formation) and faster cooling rates (to avoid Mn diffusion during cooling). Therefore, we intercritically annealed at 600°C for 15 minutes and water quenched a sample, referred to as IA600$^Q$.

The phase map and the corresponding inverse pole figure (IPF) map of IA600$^Q$ is shown in Fig. 6a. The austenite phase fraction is ~ 0.04. Contrary to IA500, austenite nucleation is observed in a the LAGB in the IA600$^Q$. Figs. 6c-f and table 5 show the correlative APT/TKD results. The 1-D concentration profiles of Mn and C across the LAGBs are shown in Figs. 6e-e' respectively. Figs. 6f-f' show the integral fraction GB excess of Mn and C respectively at the LAGBs. As observed in Fig. 6e', for the LAGB$_2$, there is a peak shift of C segregation (likely due to APT-related artifacts). For calculating the integral fraction GB excess of C, we consider that this peak corresponds to GB segregation. The integral fraction GB excess of Mn and C at the LAGBs for the IA600$^Q$ is greater compared to that observed for the IA500 (Fig. 5 and table 4). In section 4.3, we explain the temperature dependence of austenite nucleation at LAGBs based on the LAGB segregation observed in IA500 and IA600$^Q$.

*Table 5 GB excess values for the LAGBs observed in IA600$^Q$*

| Grain boundary | Misorientation (in degrees) | Mn Excess (at. % nm) | C Excess (at. % nm) | Mn in ferrite (at. %) | C in ferrite (at. %) |
|---|---|---|---|---|---|
| LAGB$_1$ | 4.5 | 25.74 | 1.71 | 8.00 | 0.38 |
| LAGB$_2$ | 5.5 | 14.72 | 2.51 | 8.39 | 0.39 |

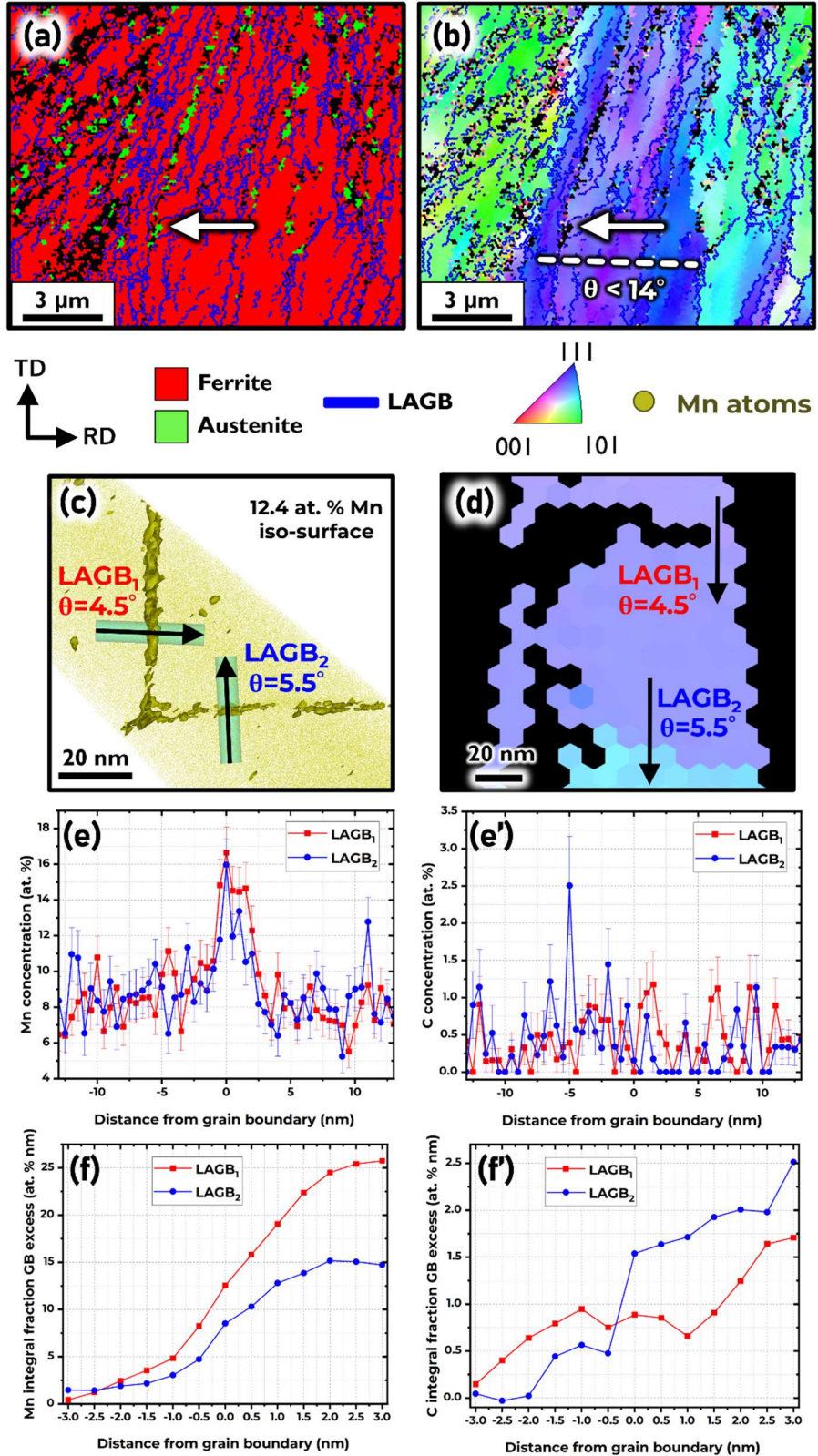

*Figure 6 (a) Phase map and (b) inverse pole figure (IPF) map of the IA600$^Q$ illustrating austenite nucleation at a LAGB. The misorientation of all the GBs within the dashed line in (b) is lower than 14°. (c-d) Correlative TKD/APT study of two LAGBs for the IA600$^Q$. (e-e') One dimensional (1-D) concentration profiles of Mn and C across the LAGBs. (f-f') Integral fraction GB excess of Mn and C at the LAGBs.*

# 4. Discussion

Previously, Nakada et al. [46] reported the temperature dependence of austenite nucleation in an ultralow carbon 13%Cr–6%Ni martensitic steel. While austenite formed at the lath boundaries at 600°C, austenite nucleation was observed at the prior austenite grain boundaries (PAGBs) at 680°C [46]. The change in free energy per unit volume was considered the same for austenite nucleation at both lath boundaries and PAGBs [46]. Using finite element analysis, the temperature-dependent austenite nucleation was explained based on the elastic strain energy and interfacial energy [46]. Lath boundaries are LAGBs with misorientation typically less than 4° [47,48]. In the 10Mn-0.05C-1.5Al (wt. %) medium Mn steel, we observed an opposite temperature dependence. At a lower annealing temperature (500°C), austenite nucleates predominantly at the HAGBs, and increasing the temperature (to 600°C) results in austenite nucleation at both LAGBs and HAGBs. Note that since the parent microstructure is 50% cold rolled martensite, the hierarchal martensitic structure [49] (such as the presence of laths, blocks, packets, and PAGBs) is no longer retained.

We observed the GB misorientation dependence of both segregation and heterogeneous austenite nucleation. We have established in section 3 that the width of GB segregation is ~1 nm. Thus, GB segregation alters both the interface energy of the GB and the local chemical driving force (section 4.1) for austenite nucleation. Hence, to clarify the role of GB misorientation-dependent segregation, we evaluate the critical radius (r*) and activation energy barrier ($\Delta G^*_{het}$) for both HAGBs and LAGBs based on classical nucleation theory (section 4.2). It is particularly relevant since LAGBs account for ~91 % of the GBs in cold-rolled martensite before intercritical annealing (supplementary S4). Finally, we explain the relation between GB segregation and the temperature dependence of austenite nucleation (section 4.3).

## 4.1. Estimating the driving force for austenite nucleation at the grain boundary

Earlier in sections 3.3 and 3.4, we demonstrated that the GB segregation width is ~1 nm and that is not limited only to GB plane. This understanding is particularly important because the GB segregation not only influences the GB energy, but also results in the change in $\Delta G_v$, the free energy per unit volume corresponding to the phase transformation. Due to the limits of the special resolution of APT (section 1 and section 2.5), we cannot experimentally determine the precise GB segregation profile. It prevents us from determining the exact chemical potential of the components ($\mu^{GB}_{Mn}$, $\mu^{GB}_C, \mu^{GB}_{Al}, and\ \mu^{GB}_{Fe}$) at the GB. To the best of authors knowledge, no thermodynamic framework exists that can account for the GB segregation width while calculating the localized driving force for nucleation; and it is beyond the scope of the present work. Therefore, to estimate $\Delta G_v$, we make a simple approximation in the current work.

Figs. 7a-c show schematically the Gibbs energy of the system vs. Mn mole fraction. $X^\alpha_{Mn}$ and $X^\gamma_{Mn}$, refer to the equilibrium Mn mole fractions of the ferrite and austenite, respectively. For the sake of simplicity, the tie line has been shown to be horizontal in the schematic in Fig. 7a-c. $\Delta G_{massive}$ can be defined as G$^\gamma$ - G$^\alpha$, which represents the driving force for a massive transformation (i. e. with no change in Mn concentration) from α to γ (schematic in Fig. 7d). The critical Mn mole fraction at which the G$^\alpha$ intersects with G$^\gamma$ is termed $X^{Crit}_{Mn}$. In other words, at $X^{Crit}_{Mn}$, $\Delta G_{massive}$ is zero. When $X_{Mn} \gg X^{Crit}_{Mn}$ (Fig. 7c), the free energy of the austenite is significantly lower than that of ferrite. This results in the $\Delta G_{massive} \cong \Delta G_{eq}$ and this understanding can also be extended to C; wherein increased C content increases the tendency for massive phase transformation (Fig. 7f).

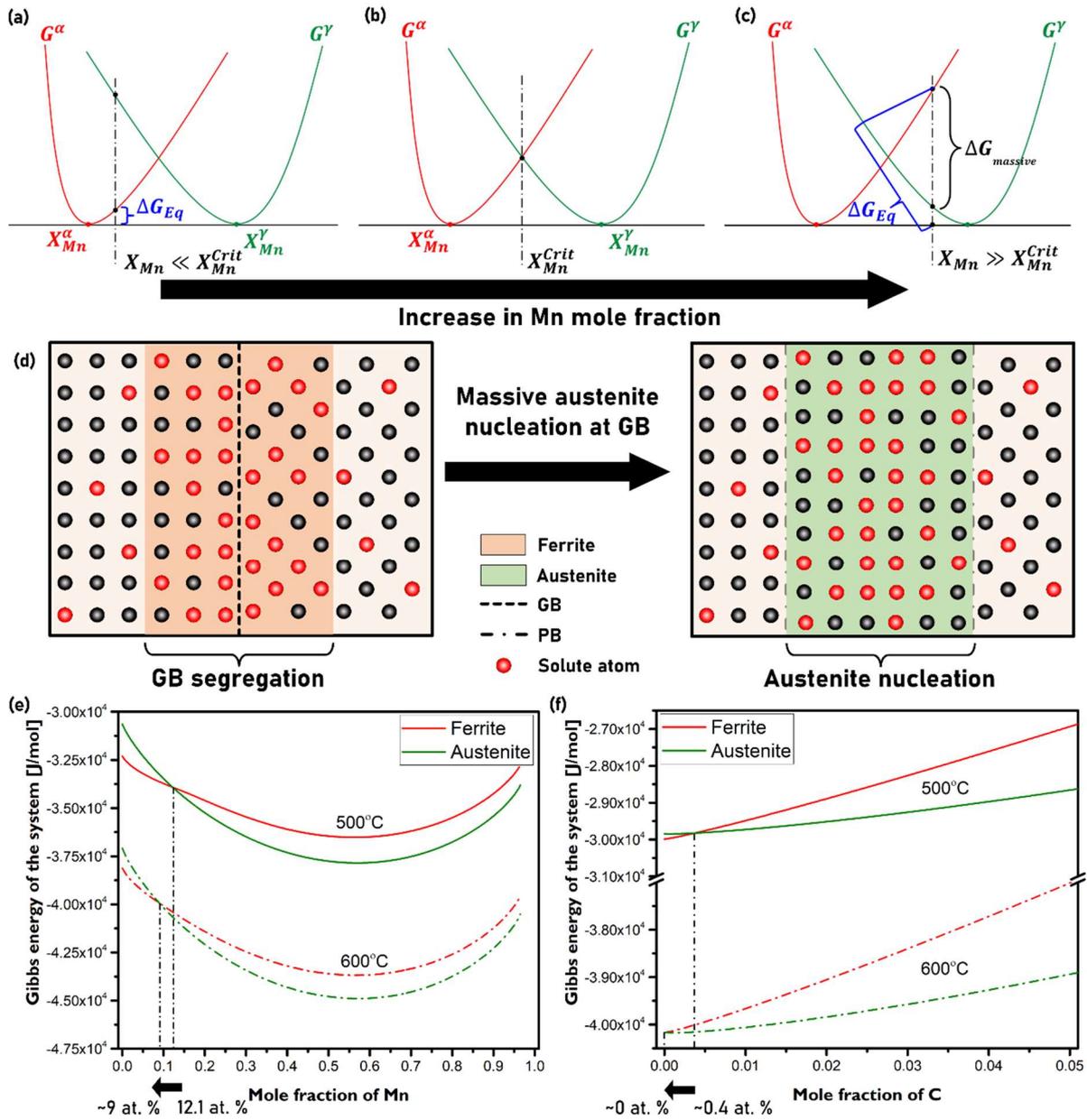

*Figure 7 (a-c) Schematic of Gibbs energy of the system with an increase in the Mn mole fraction. (d) Schematic illustrating the massive austenite nucleation; a simplistic approximation used to estimate the driving force. Thermodynamic calculations (obtained from Thermocalc) of the (e) Gibbs energy of the system vs. Mn concentration for different intercritical annealing temperatures. (f) Gibbs energy of the system vs. C concentration for different intercritical annealing temperatures*

$\Delta G_v \cong \Delta G_{massive}$ is a reasonable approximation for the localized driving force for instantaneous austenite nucleation at the GB. The physical interpretation of this approximation implies that an increase in the local Mn and C composition at the GB for a width of ~1 nm, can lead to local massive austenite nucleation at the GB (a reconstructive ("civilian") transformation) [6]. Massive transformation is 'a diffusional nucleation and growth process in which the product phase has a different crystal structure from but the same composition as the matrix phase' (Aaronson [50]). Thus, the massive transformation should not be confused with displacive ("military") diffusionless austenite reversion reported for fast heating rates [51]. Herein, the approximation of massive transformation at the GB would be a consequence of solute diffusion facilitated GB segregation. $\Delta G_{massive}$ ($G^\gamma$ - $G^\alpha$) is calculated by approximating the GB segregation excess across the GB as a 1 nm wide bulk-like BCC region with uniform composition (section 3.4 and supplementary S2). Note that this is only a simplistic estimation aimed at unravelling the effect of GB segregation on the critical radius and activation energy barrier for confined austenite nucleation at the GB.

## 4.2. Effect of segregation on the critical radius and activation energy barrier for austenite nucleation

Schematics in Figs. 8a, and 8c show austenite nucleation at the ferrite GB. Fig. 8b illustrates austenite nucleation at a triple junction. While Fig. 8a shows the scenario wherein both the austenite-ferrite phase boundaries (PBs) are incoherent, Fig. 8c corresponds to a case where one of the austenite-ferrite interfaces is coherent.

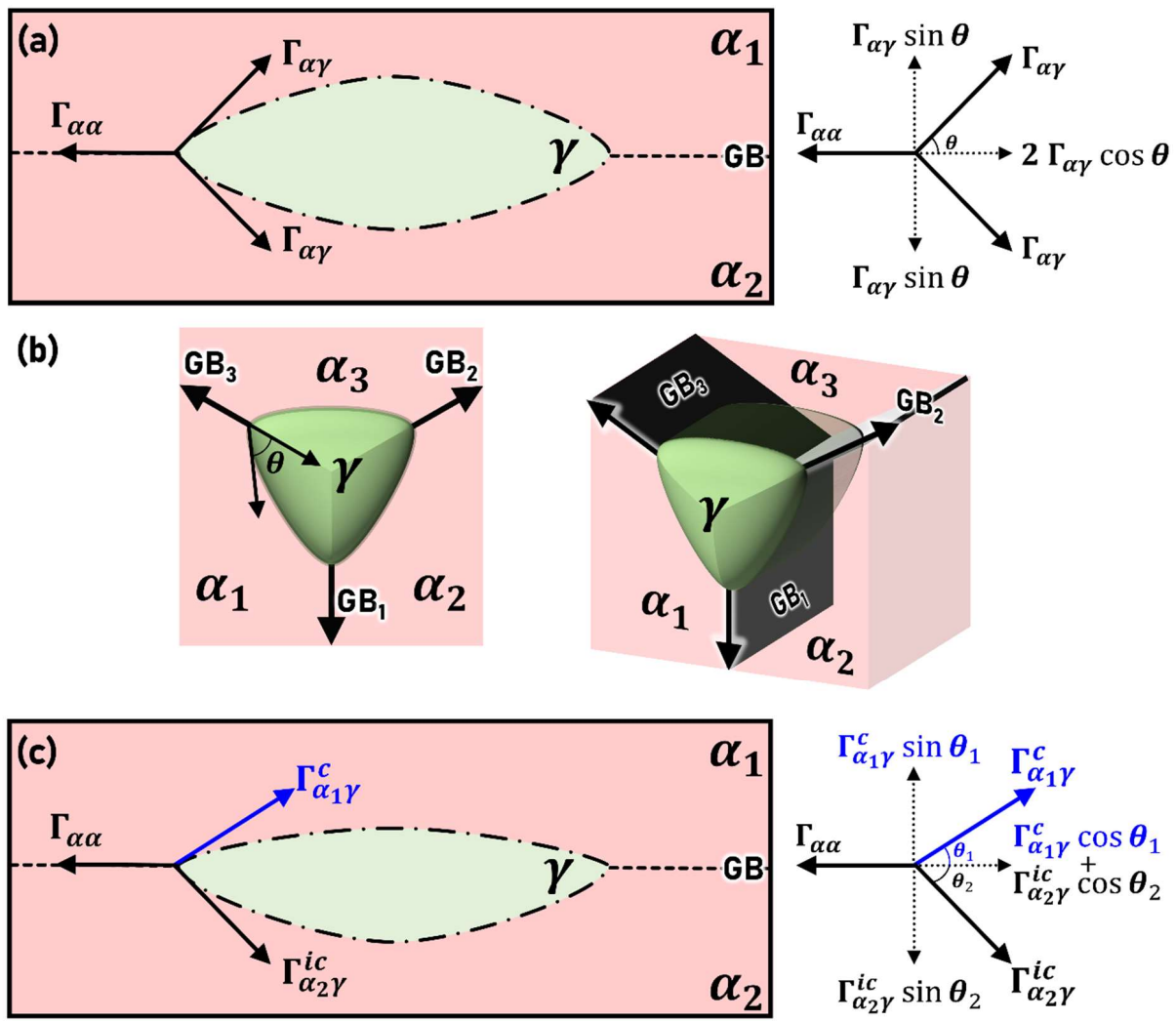

*Figure 8 Schematic diagram of austenite nucleation at the (a) ferrite grain boundary wherein both the phase boundaries are incoherent, (b) triple junction, and (c) ferrite grain boundary wherein one of the phase boundaries is coherent.*

$\Delta G_{het}$, the total free energy change for heterogeneous nucleation at a grain boundary can be given by

$$\Delta G_{het} = V\Delta G_v + A_{\alpha\gamma}\Gamma_{\alpha\gamma} - A_{\alpha\alpha}\Gamma_{\alpha\alpha} \qquad 4$$

Wherein,

$\Delta G_v$ is the change in free energy per unit volume
V is the volume of the embryo
$A_{\alpha\gamma}$ is the area of the $\alpha/\gamma$ interface of energy $\Gamma_{\alpha\gamma}$ created
$A_{\alpha\alpha}$ is the area of the $\alpha/\alpha$ grain boundary of energy $\Gamma_{\alpha\alpha}$ eliminated

$S(\theta)$ is the shape factor for heterogeneous nucleation given by

$$S(\theta) = \frac{1}{2}(2 + \cos\theta)^2(1 - \cos\theta)^2 \qquad 5.1$$

Where (Fig. 10a),
$$k = \cos\theta = \frac{\Gamma_{\alpha\alpha}}{2\Gamma_{\alpha\gamma}} \qquad 5.2$$

$$\Delta G_{het} = \Delta G_{hom} S(\theta) \qquad 5.3$$

The dislocation density in a 50% cold-rolled martensite was reported to be ~1.27 x 10$^{16}$ m$^{-2}$ [52], which is the starting microstructure before intercritical annealing in the current work. Therefore, the corresponding stored deformation energy contribution ($\Delta G_{deformation}$) of the ferrite also needs to be considered. The freshly formed austenite is assumed to be devoid of any plastic deformation. Misfit strain energy corresponding to phase transformation is ignored in the current work. Misfit strain energy arises during phase transformation due to the difference in volumes between the parent and the nucleated phase. Ferrite to austenite phase transformation is associated with ~4% volume contraction [53], which can give rise to the misfit strain energy. However, in our earlier work, after the intercritical annealing, dislocations terminating at the austenite-ferrite phase boundary were observed [4]. The presence of misfit dislocations at the interface causes strain relaxation. Additionally, experimental evidence for grain boundary migration was also presented, which can further contribute to the reduction of misfit strain energy [4]. The $\Delta G_{deformation}$ can be calculated by equation 6, wherein, ρ is the dislocation density (1.27 x 10$^{16}$ m$^{-2}$), b is 2.48 Å for ferrite, and G is the shear modulus (100 GPa) [54]. The $\Delta G_{deformation}$ is -0.03905 GJm$^{-3}$.

In the current work, we do not consider the contribution of elastic strain energy to the change in free energy during phase transformation. Nakada et al. [46] reported that the effect of elastic strain energy is critical during the martensite to austenite transformation. However, the calculations of Nakada et al. [46] have been performed with the assumption that the elastic transformation strain is never accommodated by plastic deformation and atomic diffusion at the interface. Contrary to this assumption, in the medium Mn steel studied in the current work, it has been observed earlier that both atomic diffusion and accommodation by plastic deformation (by misfit dislocations) occur at 500°C during nucleation [4]. Owing to these complexities, the calculation of the exact elastic strain contribution is beyond the scope of the current work. Consequently, we underestimate the $\Delta G_{het}$.

$$\Delta G_{deformation} = \left(G^{\gamma}_{deformation} - G^{\alpha}_{deformation}\right) = -\frac{\rho G b^2}{2} \qquad 6$$

Assuming a local massive austenite nucleation, the $\Delta G_{chem}$ is given by equation 7.

$$\Delta G_{chem} = G_v^{\gamma} - G_v^{\alpha} \qquad 7$$

The $\Delta G_v$ is the sum of $\Delta G_{chem}$ and $\Delta G_{deformation}$. Thus, the modified $\Delta G_{het}$ is given by equation 8.

$$\Delta G_{het} = V(\Delta G_{chem} + \Delta G_{deformation}) + A_{\alpha\gamma}\Gamma_{\alpha\gamma} - A_{\alpha\alpha}\Gamma_{\alpha\alpha} \qquad 8$$

For heterogeneous nucleation at grain boundary with dihedral angle $\boldsymbol{\theta}$ ($\boldsymbol{k = \cos\theta}$), and radius of curvature r [55,56];

$$A_{\alpha\alpha} = ar^2; \text{ where } a = \pi(1-k^2) \qquad 9.1$$

$$A_{\alpha\gamma} = br^2; \text{ where } b = 4\pi(1-k^2) \qquad 9.2$$

$$V = Cr^3; \text{ where } C = \frac{2\pi}{3}(2 - 3k + k^2) \qquad 9.3$$

For heterogeneous nucleation at a triple junction with dihedral angle $\theta$ ($k = \cos\theta$), and radius of curvature r [55,56];

$$A_{\alpha\alpha} = ar^2; \text{ where } a = 3\beta(1-k^2) - k\sqrt{3-4k^2} \qquad 10.1$$

$$A_{\alpha\gamma} = br^2; \text{ where } b = 12\left(\frac{\pi}{2} - \alpha - k\beta\right) \qquad 10.2$$

$$V = Cr^3; \text{ where } C = 2\left[\pi - 2\alpha + \frac{k^2}{3}(\sqrt{3-4k^2}) - k\beta(3-k^2)\right] \qquad 10.3$$

$$\alpha = \sin^{-1}\frac{1}{2\sqrt{1-k^2}} \qquad 10.4$$

$$\beta = \cos^{-1}\frac{k}{\sqrt{3(1-k^2)}} \qquad 10.5$$

Based on equations 8-9, the $\Delta G_{het}$ for austenite nucleation can be obtained by equation 10.

$$\Delta G_{het} = Cr^3(\Delta G_{chem} + \Delta G_{strain}) + br^2\Gamma_{\alpha\gamma} - ar^2\Gamma_{\alpha\alpha} \qquad 11$$

The critical radius r* of the nucleus can be estimated from equation 11.

$$r^* = \left(\frac{2(a\Gamma_{\alpha\alpha} - b\Gamma_{\alpha\gamma})}{3C(\Delta G_{chem} + \Delta G_{deformation})}\right) \qquad 12$$

$\Delta G_{het}^*$ can be further lowered if the interfaces formed are coherent in nature. Low mobility is expected for the coherent phase boundary since the migration is interface-controlled [4]. Therefore, it is unlikely that both interfaces will be coherent (as shown in Fig 8c). The nucleus can now be considered as a sum of two semi-ellipsoids with different dihedral angles, $\theta_1$ and $\theta_2$. $\Gamma_{\alpha_1\gamma}^c$ and $\Gamma_{\alpha_2\gamma}^{ic}$ are the interface energy of the coherent and incoherent phase boundaries, respectively. The dihedral angles can be obtained based on equilibrium consideration, as shown in equations 13.1-13.2. The current work, however, is restricted to the case of both phase boundaries being coherent.

$$\Gamma_{\alpha_1\gamma}^c \cos\theta_1 + \Gamma_{\alpha_2\gamma}^{ic} \cos\theta_2 = \Gamma_{\alpha\alpha} \qquad 13.1$$

$$\Gamma^c_{\alpha_1\gamma} \sin\theta_1 = \Gamma^{ic}_{\alpha_2\gamma} \sin\theta_2 \qquad 13.2$$

In the classical nucleation theory, for a given transformation temperature, $\Delta G_v$ is assumed to be the same for both homogenous and heterogeneous nucleation. The effectiveness of HAGB to nucleate is thereby ascribed to a reduction in the activation energy due to the shape factor $S(\theta)$ (equation 5). However, it is essential to note that the grain boundary segregation implies a change in the $\Delta G_v$ along with a change in the $\Gamma_{\alpha\alpha}$. Increased segregation, as shown in table 4, not only increases the driving force $\Delta G_v$, but also translates into a reduction in the critical radius needed for the nucleation of austenite (equation 10).

In table 6 (and Fig. 9), the critical radius for austenite nucleation (r*) and $\Delta G^*_{het}$ is summarized for different scenarios for the intercritical annealing temperature of 500°C. For calculations in table 6, the GB energy for HAGB and LAGB are 1.2 J/m² and 0.45 J/m², respectively (section 3.3). GB segregation reduces the GB energy, as shown earlier in section 3.3. The GB energy post-segregation is reduced by a factor of 0.55 (Table 3). Chemical driving force values for HAGB and LAGB have been adapted from table 4. The driving force for LAGB is the average driving force value reported for all the LAGBs in table 4. The interface energy values are assumed to be 0.4 J/m² for the coherent and 1.3 J/m² for the incoherent phase boundaries [30]. The Total driving force (obtained from equation 8) considers both the chemical and deformation energy contributions. At 500°C, when GB segregation is not considered, $G^\alpha_v < G^\gamma_v$ and therefore, the chemical driving force of the nominal composition is estimated by $\Delta G_{eq} = G^\alpha_v - G^{system}_{equilibrium}$. For a given composition, chemical driving force estimated by approximating to a massive transformation would be lower than the equilibrium driving force (Fig. 7c).

*Table 6 Reduction in critical radius r* for austenite nucleation at 500°C due to GB segregation.*

| GB Type | GB energy (J/m²) | GB segregation | GB energy (after seg.) (J/m²) | Interface energy (J/m²) | Chemical driving force (GJ/m³) | Total driving force (GJ/m³) | Critical radius (nm) | $\Delta G^*_{het}$ (10⁻¹⁸) J |
|---|---|---|---|---|---|---|---|---|
| **Grain boundary** | | | | | | | | |
| HAGB | 1.2 | x | x | 1.3 | -0.033 | -0.07205 | 52.74 | 9169.63 |
| | | ✓ | 0.66 | 1.3 | -0.27 | -0.30905 | 10.55 | 494.92 |
| | | ✓ | 0.66 | 0.4 | -0.27 | -0.30905 | 4.72 | 7.02 |
| LAGB | 0.45 | x | x | 1.3 | -0.033 | -0.07205 | 42.33 | 8646.58 |
| | | ✓ | 0.2475 | 1.3 | -0.103 | -0.14205 | 20.05 | 2065.11 |
| | | ✓ | 0.2475 | 0.4 | -0.103 | -0.14205 | 7.37 | 69.65 |
| **Triple Junction** | | | | | | | | |
| HAGB | 1.2 | x | x | 1.3 | -0.033 | -0.07205 | 36.0863 | 1422.19 |
| | | ✓ | 0.66 | 1.3 | -0.27 | -0.30905 | 8.41 | 190.18 |
| | | ✓ | 0.66 | 0.4 | -0.27 | -0.30905 | 2.59 | 0.01 |
| LAGB | 0.45 | x | x | 1.3 | -0.033 | -0.07205 | 36.0861 | 4531.42 |
| | | ✓ | 0.2475 | 1.3 | -0.103 | -0.14205 | 18.30 | 1448.39 |
| | | ✓ | 0.2475 | 0.4 | -0.103 | -0.14205 | 5.63 | 21.42 |

Based on table 6 and Fig. 9, the following inferences can be made:

1. GB segregation results in the increase of the driving force, thereby resulting in a decrease of the critical radius for austenite nucleation (r*). The greater the segregation, the higher the driving force for austenite nucleation and the lower the critical radius r*. This explains the greater propensity for austenite nucleation at HAGBs.

2. The critical radius for austenite nucleation (r*) is lowered due to the formation of coherent phase boundaries with low interface energy (0.4 J/m²).

3. In the case of the absence of segregation, the critical radius for austenite nucleation (r*) is nearly the same when the triple junction is formed either from HAGBs or LAGBs. However, $\Delta G^*_{het}$ ($\Delta G_{het}$ corresponding to r*) is significantly lower for the triple junction made of HAGBs.

4. A segregated HAGB is more favorable for austenite nucleation (both lower critical radius r* and $\Delta G^*_{het}$) when compared to a triple junction formed by segregated LAGBs. This reinforces the importance of segregation, as discussed earlier.

5. As mentioned earlier, the elastic strain energy was not included in the calculations. Therefore, the critical radius (r*) shown in table 6 is the upper bound value. The actual critical radius of austenite nucleation (r*) is likely to be much lower, closer to the 1 nm segregation width, as discussed in section 3.3.

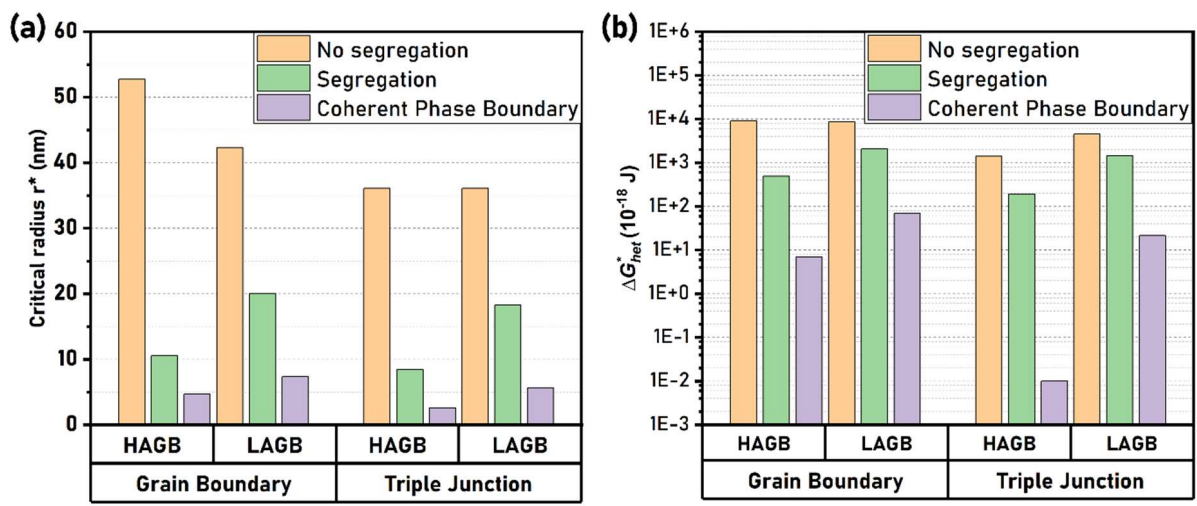

*Figure 9 Critical radius for austenite nucleation (r*) and $\Delta G^*_{he}$ for different scenarios shown in table 5. The coherent phase boundary corresponds to scenario with both segregation and coherent austenite-ferrite interfaces present.*

### 4.3. Temperature dependence of GB segregation driven austenite nucleation

As discussed in section 4.2, the GB segregation not only contributes to the driving force but can also help reduce the critical radius r*. In the current section we further demonstrate that for a given LAGB chemistry, an increase in temperature increases the chemical driving force, thereby enhancing the austenite nucleation tendency. Figs. 7e-f show the temperature and compositional (Mn and C, respectively) dependence of Gibbs free energy for austenite and ferrite. Thermo-Calc calculations suggest (Figs. 7e-f) the $X^{Crit}_{Mn}$ decreases from 12.1 at.% to 9 at.% with an increase in temperature from 500°C to 600°C. Similarly, the critical C concentration $X^{Crit}_C$ decreases from 0.4 at.% to ~0 at.% with the increase in the intercritical annealing temperature from 500°C to 600°C. In Fig. 11, the chemical driving force values for austenite nucleation estimated by Thermo-Calc for different Mn and C compositions at 500°C and 600°C are reported. It is observed that for lower Mn and C concentrations, the corresponding driving force is significantly greater at 600°C compared to 500°C (Fig. 11a-c). For instance, for 12.5 at. % Mn and 0.25 at. % C at 500°C, the driving force is nearly zero. However, at 600°C, the driving force is ~300 J/mol. Thus, for a given GB chemistry (for lower segregations), an increase in temperature increases the tendency of nucleation at LAGBs (Figs. 10a-c). This explains the temperature dependence of austenite nucleation at the LAGB.

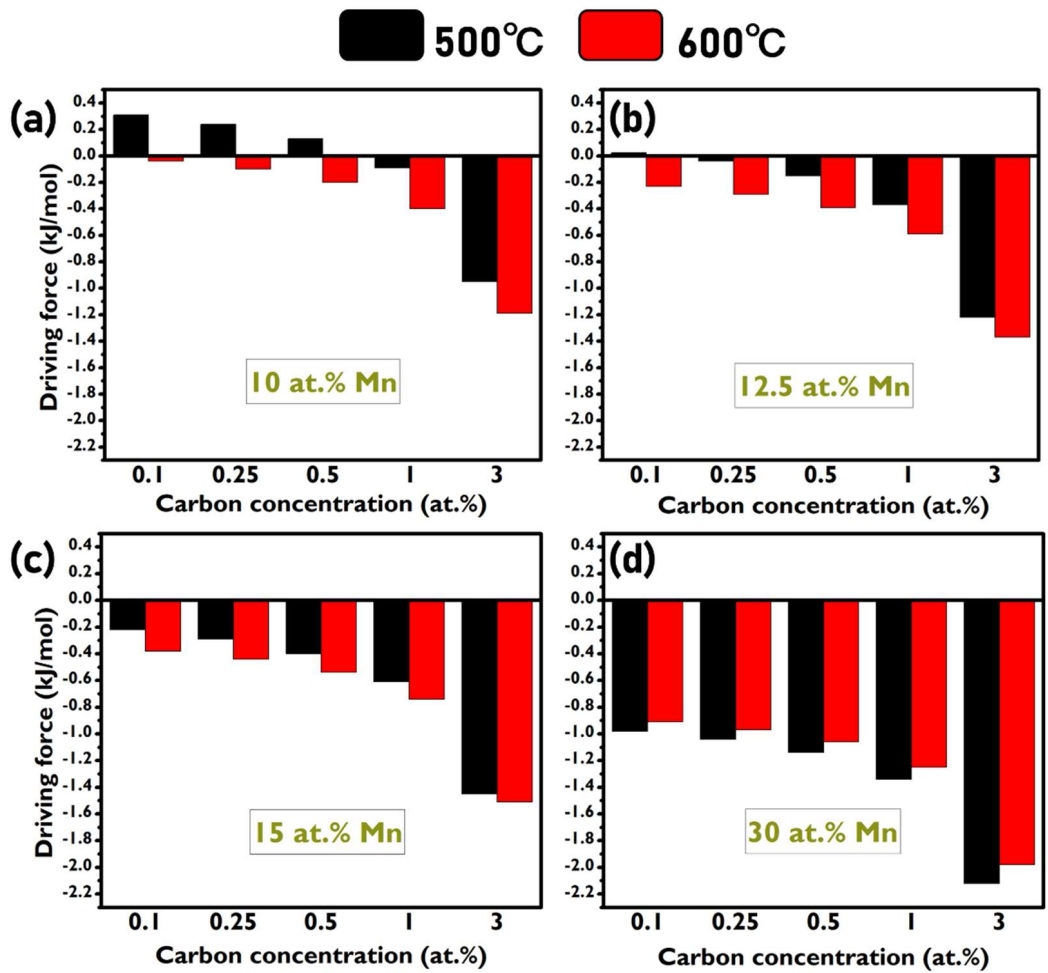

Figure 10 Chemical driving force ($\Delta G_{chem} = G_v^\gamma - G_v^\alpha$), at 500°C and 600°C for different C concentrations (0.1 at.%, 0.25 at.%, 0.5 at.%, 1 at.% and 3 at.%) for (a)10 at. % Mn (b) 12.5 at.% Mn (c) at.% Mn and (d) 30 at.% Mn

### 4.4. Effect of GB carbon depletion on austenite nucleation: Case study of Nb micro-alloying

Previously, we added 0.06 wt. % Nb to the medium Mn steel investigated in the current work and reported the effect of Nb micro-alloying on austenite nucleation and growth [57]. For an intercritical annealing temperature of 550°C, we observed that Nb microalloying reduced the quantity of austenite nucleation events by 20% (evidenced by the number density of austenite grains) [57]. It was attributed to the carbon depletion at the GBs due to NbC precipitation [57].

There are two possible scenarios for the effect of carbon GB segregation on the chemical driving force for austenite nucleation: (i) it has no influence, and (ii) carbon segregation increases the chemical driving force, promoting austenite nucleation (current work). Assuming scenario (i), GB segregation only reduces initial GB energy ($\Gamma_{\alpha\alpha}$). It follows that the GB energy of a segregation-depleted boundary is greater than the GB energy of a segregated boundary. When a GB is eliminated during heterogeneous nucleation, the higher the GB energy, the greater the driving force for phase transformation, and the lower the activation energy barrier (section 1). If scenario (i) were to be true, the depletion in carbon segregation due to Nb micro-alloying should have enhanced austenite nucleation at the GBs. On the contrary, depletion in carbon segregation was accompanied by a 20% reduction in austenite nucleation [57]. Hence, the case study provides evidence that solute segregation at GBs influences the chemical driving force. In the case of medium Mn steels, C and Mn segregation at the GBs promotes austenite nucleation.

## 5. Conclusions

In the present work, we explain the effect of grain boundary misorientation-dependent solute segregation on the heterogeneous austenite nucleation in medium Mn steels. The conclusions are summarized herein:

1. DFT studies revealed that at 500°C the maximum Mn segregation tendency was not limited to the GB plane but rather can occur at the plane(s) adjacent to the GB. In other words, the segregation is not confined to the GB plane and the solute also enriches the bulk lattice adjacent to the GB plane (~ 0.5 nm on either side of the GB). Hence, the segregation values obtained from APT results should not be attributed to the GB plane alone.

2. We employed a correlative TKD/APT approach to study the structural and chemical information of ferrite GBs. By calculating the integral fraction GB excess, we accounted for both the inherent GB segregation width (1 nm) and inaccuracies in segregation width profiles arising due to APT-related artifacts.

3. We approximated the localized driving force to be the driving force for massive transformation at the grain boundary. Thus, the greater the segregation, the higher the chemical driving force for austenite nucleation. Significantly greater Mn and C segregation is observed at HAGB compared to the LAGB. Consequently, the chemical driving force for austenite nucleation at HAGB is higher than at LAGB. The greater the driving force for austenite nucleation, the lower the critical radius r*. This results in preferential nucleation of austenite at the HAGB.

4. The estimations showed that at 500°C, a segregated HAGB is more favourable for austenite nucleation (both lower critical radius r* and $\Delta G^*_{het}$) when compared to a triple junction formed by segregated LAGBs. This further emphasizes the importance of GB segregation in austenite nucleation.

5. Austenite nucleation at LAGBs was nearly absent for the IA500. Conversely, austenite nucleated at LAGBs for the IA600. The temperature-dependent austenite nucleation at LAGBs was explained.

## 6. Acknowledgements

This work is funded by Tata Steel, IJmuiden, Netherlands, through the IMPRS SURMAT scholarship. The authors express their gratitude to Prof. Tilmann Hickel for the discussions on DFT calculations. R. S. Varanasi thanks Dr. Reza Darvishi Kamachali and Prof. Dierk Raabe for the insightful discussions. The authors are grateful to Uwe Tezins, Christian Broß, and Andreas Sturm for their support to the FIB and APT facilities at MPIE.## 7. Declaration of Competing Interest

The authors declare that they have no known competing financial interests or personal relationships that could have appeared to influence the work reported in this paper.

# Supplementary

# Temperature and misorientation-dependent austenite nucleation at ferrite grain boundaries in a medium manganese steel: role of misorientation-dependent grain boundary segregation

*Rama Srinivas Varanasi[1,2], #Osamu Waseda[1], Faisal Waqar Syed[1], Prithiv Thoudden-Sukumar[1], Baptiste Gault[1,3], Jörg Neugebauer[1], †Dirk Ponge[1]

[1]Max-Planck-Institut for Sustainable Materials, Max-Planck-Straβe 1, 40237, Düsseldorf, Germany

[2] Institute for Materials Research, Tohoku University, 2-1-1 Katahira, Aoba-ku, Sendai, 980-8577 Japan

[3]Department of Materials, Royal School of Mines, Imperial College, Prince Consort Road, London SW7 2BP, United Kingdom.

**Corresponding authors**

*Rama Srinivas Varanasi: rama.varanasi@tohoku.ac.jp

#Osamu Waseda: o.waseda@mpie.de

†Dirk Ponge: d.ponge@mpie.de

## S1.    Heterogenous nucleation at grain boundary

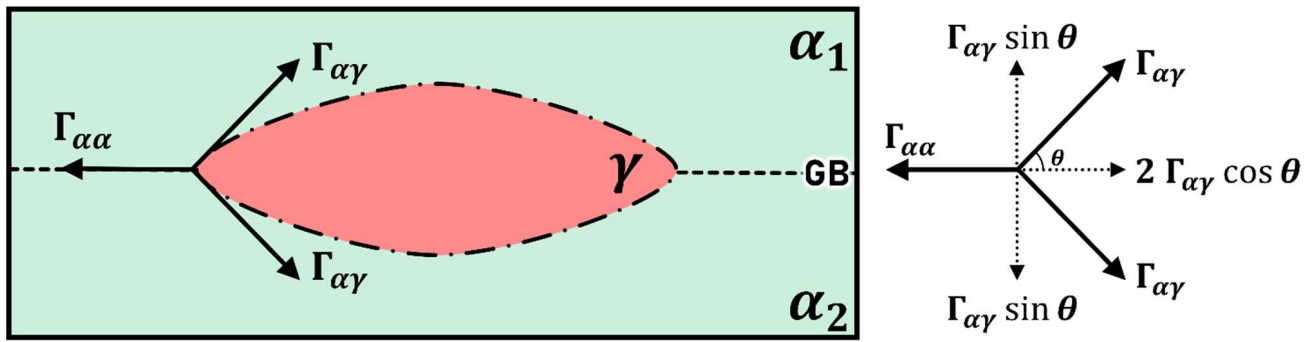

**Figure S1** Schematic of heterogeneous nucleation of austenite (γ) at a ferrite (α) grain boundary (GB).

Fig. S1 shows a schematic of heterogeneous nucleation at grain boundary (GB). Based on Fig. SI

$$\cos\theta = \frac{\Gamma_{\alpha\alpha}}{2\,\Gamma_{\alpha\gamma}} = A \tag{1}$$

Based on the classical nucleation theory:

$$\frac{\Delta G^*_{het}}{\Delta G^*_{hom}} = S(\theta) = \frac{1}{2}(2+A)(1-A)^2 \tag{2}$$

Since, $0 \leq \theta \leq \frac{\pi}{2}$ (Fig. S1), $1 \geq A \geq 0$. Fig. S2 illustrates the change in $S(\theta)$ with respect to A i.e., $\cos\theta$. The greater the value of A, the lower the value of $S(\theta)$. In other words, for a given value of $\Gamma_{\alpha\gamma}$, the greater the energy of the grain boundary ($\Gamma_{\alpha\alpha}$), lower is the activation energy barrier ($\Delta G^*_{het}$) in a heterogeneous nucleation.

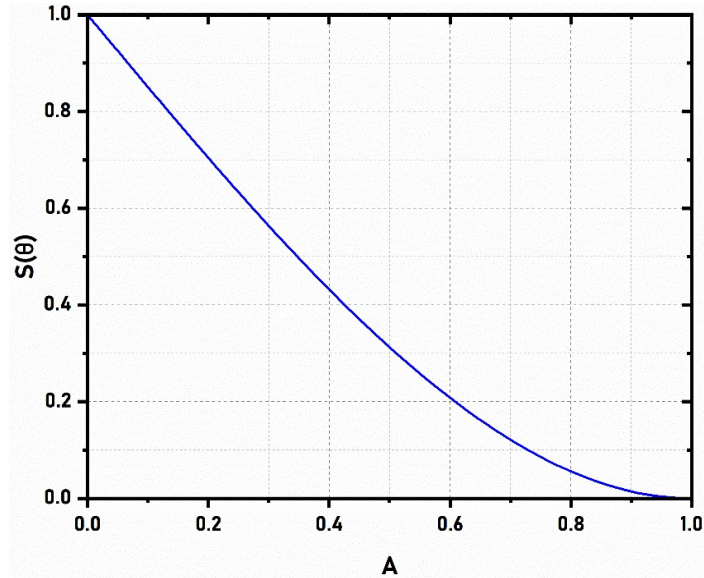

**Figure S2** Plot of the change in $S(\theta)$ with respect to change in $\cos\theta$ i.e., (A).

## S2. Integral fraction grain boundary excess calculation

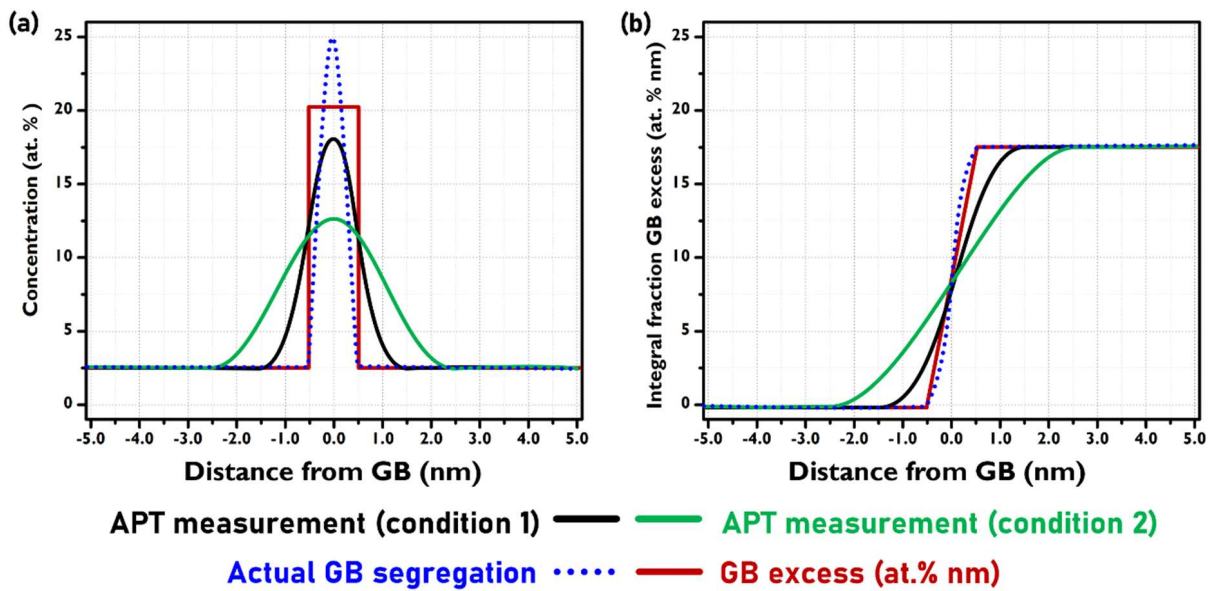

**Figure S3** *Schematic of the integral GB excess calculation adapted from Maugis et al. [41]. (a) The blue dotted curve represents the actual GB segregation profile. For the given GB, black and green curves are experimentally segregation profiles corresponding to different APT measurement conditions. The red curve represents the physical significance of GB excess reported as at. % nm: approximation of GB segregation to 1 nm wide uniform segregation profile across the GB. (b) Maugis et al. [1] proposed that the integral GB excess value obtained from the APT measurements will be similar to the actual segregation profile.*

The schematic in Fig. S3a shows different GB segregation profiles across the grain boundary, and Fig. S3b shows the corresponding integral fraction of GB excess across the grain boundary. In Fig. S3, the blue dotted curve represents the actual segregation profile. However, the observed concentration profiles for a given GB can vary from measurement to measurement (black and green curves in Fig. S3) due to evaporation artifacts associated with atom probe measurements. Maugis et al. [1] proposed that the integral GB excess values for the measured values should be similar to that of the actual segregation, as shown in Fig. S3b. The integral GB excess is represented in the units at. % nm. In other words, it is the GB excess associated with a uniform GB segregation across 1 nm (red curve in Fig. S3a-b). For example, if the bulk composition is 2.5 at. % (Fig. S3a), and the integral GB excess is 17.5 at. % nm (Fig. S3b), the solute concertation can be approximated to 20 at. % over a GB width of 1 nm (red line in Fig. S3a). In the present work, we refer to this as 'adjusted GB segregation value' (red curve in Fig. S3a).

## S3. Spinodal decomposition at the HAGB

We observe spinodal decomposition at the GBs. Fig. S3 illustrates the spinodal decomposition at a HAGB wherein we observe a Mn compositional spread between 14-21 at. %.

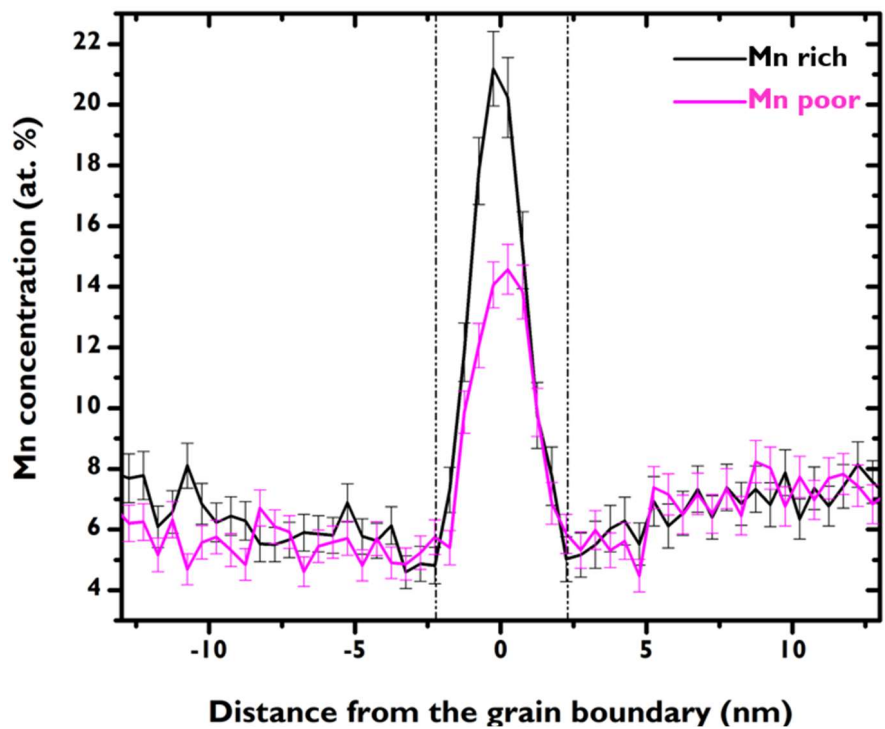

**Figure S4** *Spinodal decomposition at a HAGB wherein we observe Mn compositional spread between 14-21 at. %.*

## S4. GB misorientation angle distribution

Fig. S4 shows a grain boundary misorientation angle distribution for cold-rolled martensite before intercritical annealing, wherein a bimodal distribution is observed.

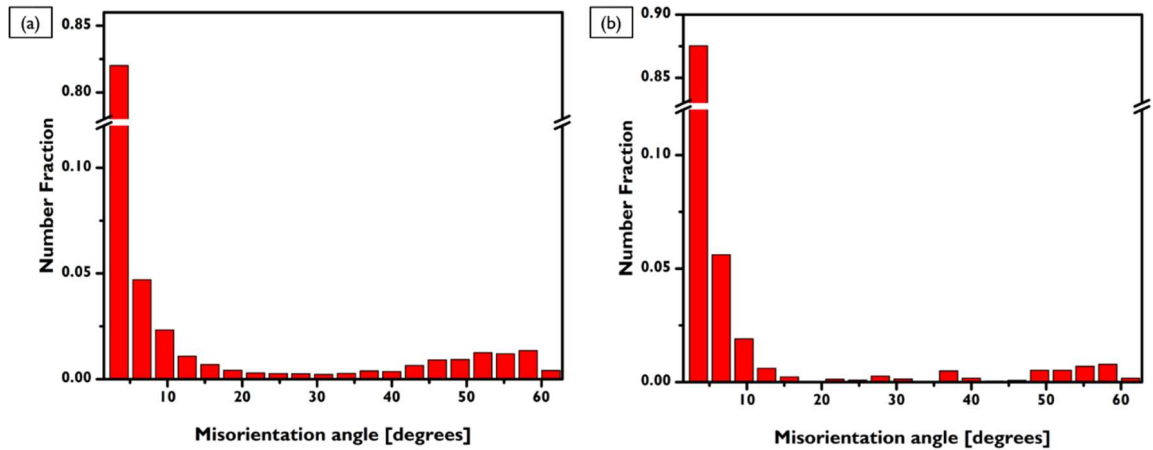

**Figure S4** *Misorientation angle distribution for the 50% cold-rolled medium manganese steel (a) prior to annealing (b) after intercritical annealing at 600°C for hours.*

## References

[1] P. Maugis, K. Hoummada, A methodology for the measurement of the interfacial excess of solute at a grain boundary, Scripta Materialia 120 (2016) 90–93. https://doi.org/10.1016/j.scriptamat.2016.04.005.